\documentclass[twocolumn,showpacs,preprintnumbers]{revtex4}
\usepackage{amsmath}
\usepackage{amssymb}
\usepackage{graphicx}
\usepackage{bm}
\begin{document}
\title{Exotic break-up modes in heavy ion reactions at low energies}
\author{C. Rizzo$^{1}$, M. Colonna$^{1}$, V. Baran$^{2}$ and M. Di Toro$^{1,3}$} 
\affiliation{
        $^1$INFN-LNS, Laboratori Nazionali del Sud, 95123 Catania, Italy\\
        $^2$ Faculty of Physics, University of Bucharest, Romania\\
	$^3$Physics and Astronomy Dept. University of Catania, Italy\\}

\begin{abstract}
New reaction mechanisms occurring in heavy ion collisions at low energy
(10- 30 MeV/A) are investigated within the Stochastic Mean Field 
model. We concentrate on the analysis of ternary breakup events, of
dynamical origin, occurring 
in semi-central reactions, where the formation of excited systems in various 
conditions of shape and angular momentum is observed.  
We show how this fragmentation mode, which can be
considered as a precursor of the neck emission observed at higher
beam energies, emerges from the combined action of surface (neck) instabilities and
angular momentum effects. 
Interesting perspectives are opening towards the investigation of this mechanism
in neutron-rich (or exotic) systems, with the possibility to access
information on the low-density behavior of the nuclear symmetry energy. 



\end{abstract}
\maketitle

\section{Introduction}

At energies from just above the Coulomb barrier up to the Fermi regime, heavy ion collisions are
largely dominated by dissipative one-body reaction mechanisms.
According to the selected impact parameter window, the reaction dynamics ranges mainly from 
(incomplete) fusion to binary channels, associated with deep-inelastic and/or dynamical fission 
processes \cite{Laut06}.
Owing to the complex mean-field dynamics, 
the system may explore, along the separation path, rather extreme conditions with respect to
shape and angular momentum, which may induce large fluctuations in the exit channel.   
The mechanisms governing the transition between fusionlike and binarylike processes 
represent  a long-lasting
subject of investigation \cite{Aw84,Pen90,Denis}. In particular, much attention has been devoted to
the possible origin of the large variances 
observed for the reaction products \cite{Aw84,Pen90,Amo09,Col95,Cedric}. 

The very dissipative dynamics may also lead, especially in collisions
between medium-heavy systems, to the development of new modes of reseparation, such as dynamical 
ternary or quaternary breaking.  The occurrence of this reaction outcome,
characterized by  massive fragments nearly aligned along a common separation axis, 
is well documented \cite{Gla,Stef} and has been recently reported in  $^{197}$Au + $^{197}$Au
collisions at 15 and 23 MeV/nucleon \cite{Isa,Cap}. 
The intricate neck dynamics and the formation of rather elongated Projectile-like (PLF)
and Target-like (TLF) fragments, in semi-peripheral reactions, 
could be at the origin of further rupture steps, 
leading to the observation of three or four fragments in the exit channel.
Large scale quantum and thermal fluctuations of the nuclear mean-field
are expexted to play a crucial role in this process, thus opening the possibility to learn about
important ingredients of the nuclear effective interaction \cite{Cedric}.

 Another interesting aspect is the possibility to consider the ternary breakup
channel as a precursor of the neck fragmentation mechanism observed
at higher beam energies. Indeed, in semi-peripheral collisions
at 30-50 MeV/nucleon, the low-density neck region which develops between the two
reaction partners becomes so pronounced that small
fragments are directly emitted on very short time scales, 
with larger relative velocity with respect to PLF and TLF \cite{baran2004,Def,Def_new,Sylvie,Silvia}.

In reactions involving neutron-rich (or even exotic) nuclei, one may expect the multi-breakup
probability to be affected by the dynamics of the neutron-rich neck region,
which, in turn, can be influenced by specific properties of the nuclear
effective interaction. 
Indeed from these studies it may become possible to access information on the low-density
behavior of the isovector term of the nuclear potential \cite{Paul}
and the corresponding symmetry energy of the nuclear equation of state, on which many investigations
are concentrated nowadays, see  Ref.\cite{EPJA_book} for a recent review.

From the above discussion, it is clear 
that a detailed analysis of the mean-field dynamics, and associated shape
fluctuations and rotational effects, is crucial to investigate 
the competition between reaction
mechanisms, 
as well as the nature of new exotic reseparation modes,
in low energy dissipative collisions.
In this paper, we undertake such a study in the framework of the Stochastic Mean Field (SMF) 
transport model, which has been shown to provide a good description of
mean-field dynamics, as well as of the effect of two-body fluctuations and correlations \cite{rep,rep1,Leonid,Rizzo_2011}.
We will investigate semi-peripheral heavy ion collisions in the beam energy 
range of 10-30 MeV/nucleon,
focusing on the possible occurrence of ternary breaking and on the
features of the associated reaction products. 
This may open a novel understanding of the transition path from
deep-inelastic collisions to multifragmentation processes and elucidate
the nuclear dissipation mechanisms. 

The paper is organized as it follows: In Section 2 we present the SMF transport treatment
employed to 
follow the dynamical evolution of nuclear collisions. 
Results concerning the features of multi-step breaking mechanisms
are discussed in Section 3.  Finally conclusions and perspectives
are drawn in Section 4.


\section{Dynamical description of nuclear reactions}

The description of the nuclear reaction dynamics is afforded considering, as 
a starting point, 
the Boltzmann-Langevin (BL) equation, wich defines the time evolution of
the semiclassical one-body distribution function 
$f({\bf r},{\bf p},t)$ (i.e. the semi-classical analog of the Wigner transform of
the one-body density matrix):
\begin{equation}
\frac{\partial f}{\partial t}+\frac{\bf p}{m}\frac{\partial f}{\partial {\bf r}}-
\frac{\partial U}{\partial {\bf r}}\frac{\partial f}{\partial {\bf p}}=I_{coll}[f]+\delta I[f].
\end{equation}
The coordinates of isospin are not shown for brevity.
Eq.(1) essentially describes the behavior of the system
in response to the action of the self-consistent mean-field potential $U$, whereas effects of two-body 
correlations and fluctuations are incorporated in the collision integral, $I_{coll}$, and its stochastic
part, $\delta I$ \cite{Ayik,Rizzo_BL}.
The average term $I_{coll}[f]$ takes into account the 
energy, angular and isospin dependence of free nucleon-nucleon cross sections \cite{baran2002}.

We adopt the following parametrization of the mean-field potential:
\begin{eqnarray}
U_{q}&=&A\frac{\rho}{\rho_0}+B(\frac{\rho}{\rho_0})^{\alpha+1}
+C \frac{\rho_n-\rho_p}{\rho_0}\tau_q \nonumber \\
\end{eqnarray}
where $\rho$ denotes the density, $q=n,p$ and $\tau_n=1, \tau_p=-1$.
The coefficients $A=-356~MeV$,$B=303 ~MeV$ and the exponent $\alpha=\frac{1}{6}$, 
characterizing the isoscalar part of the mean-field, are fixed requiring 
that the saturation properties of symmetric 
nuclear matter, $\rho_0=0.16 fm^{-3}$ and $E/A=-16~MeV/nucleon$,
with a compressibility of $200$$~MeV$, are reproduced. 
We notice that the considered compressibility value is favored e.g. 
from flow, monopole oscillation and multifragmentation studies \cite{rep,Borderie}. 
This choice corresponds
to a Skyrme-like effective interaction, namely $SKM^*$, for which we consider the effective
mass as being equal to the nucleon bare mass. 
As far as the isovector part of the nuclear interaction is concerned, 
we take a constant value of $C = 36~MeV$, corresponding to a linear (stiff)
behavior of the potential part of the symmetry energy, 
$C_{sym}^{pot} = 36\rho/(2\rho_0)$~\cite{baran2002,rep1}.

Within such a framework, 
the system is described in terms of the one-body distribution function $f$, but this function
may experience a stochastic evolution in response to the action of the fluctuating term $\delta I[f]$.
The Stochastic Mean Field (SMF) model, that we adopt here, represents an approximate approach to solve the BL equation, 
where phase-space fluctuations are projected in coordinate 
space \cite{TWINGO,Salvo}. 
Thus the fluctuating term $\delta I[f]$ is implemented through stochastic spatial density fluctuations.
Eq.(1) is solved numerically adopting the test particle method \cite{TWINGO}. 



It should be noticed that 
semi-classical models have been shown to work well for the description
of the approaching phase of reactions at energies just above the Coulomb barrier, leading
to the formation of composite excited systems \cite{Baran96,Papa,QMD}.
Moreover, the inclusion of fluctuations in the dynamics allows one to address mechanisms governed by the growth
of mean-field instabilities, such as surface break-up processes, occurring at low energies \cite{David,Leonid,Rizzo_2011},
 or volume (spinodal) decomposition, 
leading to multifragmentation events at Fermi energies~\cite{rep,Akira}.

The fluctuations implemented in the SMF model are essentially of thermal nature.
Indeed, in semi-classical approaches quantal fluctuations cannot be accounted for. 
However, though the amplitude of fluctuations may be underestimated,
the SMF model has been shown to  
deal properly with the development of surface instabilities (or meta-stabilities) characterizing
heavy ion collisions around 10-20 MeV/nucleon, i.e. the formation
of primary reaction products with large quadrupole and/or octupole deformation.
As shown in Refs. \cite{Leonid,Rizzo_2011}, the analysis of shape observables, such as
multipole moments,  allows one to extract valuable information about
fusion vs. break-up probabilities, in low energy semi-central reactions. 
Here we extend this kind of study to the possible occurrence of
ternary break-up events, on which recent experimental investigations
have been concentrated \cite{Isa,Cap,wilc,wilc2}.



 

\begin{figure}[t]
\includegraphics[width=9.cm]{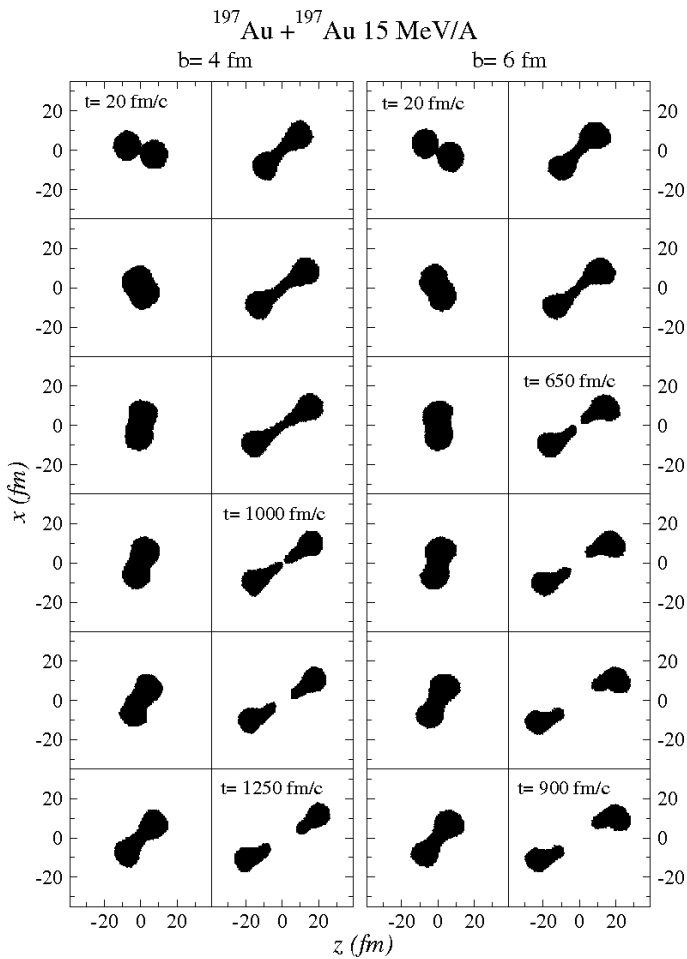}
\caption{Contour plots of the density projected on the reaction plane,
  calculated with SMF, for the reaction $^{197}$Au +
  $^{197}$Au at 15 MeV/nucleon, at several times, 
from top to bottom. 
The time instants corresponding to the splitting of the system
in PLF-TLF fragments ($t_{sep}$) and to the final calculation
time ($t_{stop}$) are indicated in the figure. 
Results corresponding to impact parameters
b = 4 fm and b = 6 fm are displayed.}
\label{contour_BGBD}
\end{figure}

\begin{figure}[t]
\includegraphics[width=9.cm]{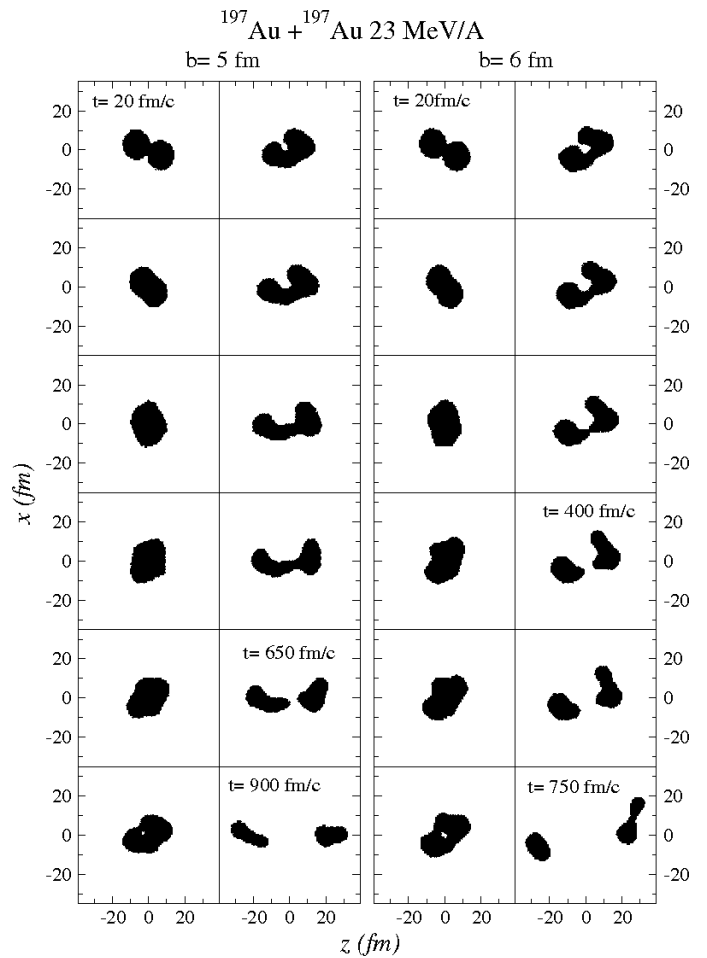}
\caption{The same as in Fig.1, for the reaction at 23 MeV/A. 
 Results corresponding to impact parameters
b = 5 fm and b = 6 fm are displayed.}
\end{figure}

\section{Results}
We focus on the study of the $^{197}$Au + $^{197}$Au reaction in the beam-energy range of 10-30 MeV/nucleon
\cite{Isa,Cap,wilc}. 
At 15 MeV/nucleon and at 
semi-central impact parameters, corresponding to strongly damped collisions \cite{wilc2}, 
the data contain predominantly binary events. However, break-up processes into three or four massive fragments of 
comparable mass are also revealed.
The ternary events are dominated by configurations where the heaviest fragment, close to the $^{197}$Au mass, 
is recognized as the remnant of the projectile (PLF) or the target (TLF), while the other two fragments, 
indicated as F1 and F2 in the analysis of Ref.\cite{wilc}, are generated by the subsequent break-up of TLF or PLF.

The observed F1 and F2 fragments exhibit similar masses. Moreover, for the considered ternary partitioning mechanism, 
the three fragments are almost aligned 
along the axis determined by the TLF (or PLF) velocity and the velocity of the reconstructed PLF (or TLF). 
We notice that the latter feature is not compatible with the scenario of a pure statistical fission, pointing to a dynamical
origin of the fragment formation mechanism. 
It is also observed that, in the majority of the events, the fragment with the largest parallel velocity (denoted
as F1 in the experimental analysis) has the smallest mass.

The experimental study of the same system at higher beam energy (23 MeV/nucleon) 
identified a larger variety of fragment sizes \cite{Cap}. 
Indeed, together with the observation of ternary events with fragments of comparable size, 
an abundant emission of Intermediate Mass Fragments (IMF), with mass less than 20, accompanied by two heavy fragments 
(TLF and PLF),
was detected in a large fraction of events. 
This observation supports the transition, when increasing the beam energy,
from dynamical ``fast fission'' processes
to the neck fragmentation mechanism, which has been widely investigated at Fermi energies \cite{baran2004,Def,Def_new,Sylvie,Silvia}. 

The aim of our theoretical analysis is to study the collision path leading to ternary break-up configurations,
to probe the possible dynamical origin and the role of mean-field instabilities
in the corresponding reaction mechanism. 
The nuclear reactions described above are investigated in the framework of the SMF model, using 100 test particles per nucleon, which ensure
an accurate description of the mean-field dynamics.  
100 events are considered for each set of macroscopic initial conditions (i.e. beam energy and impact parameter).

In Fig.1 we present density contour plots (in the reaction plane) 
obtained in one representative event of the reaction $^{197}$Au + $^{197}$Au 
at 15 MeV/nucleon, for b=4 fm (left panel) and b=6 fm (right panel).  
The impact parameter region considered in our calculations, [4-6] fm, corresponds
to rather dissipative collisions, where an intricate neck dynamics is perceived, leading eventually to the possibility
of observing multiple break-up.
Indeed, from the plots shown in Fig.1, one can already appreciate that 
the reaction mechanism is heavily dominated by the occurrence of fragment quadrupole and octupole deformations in the exit channel. 
The neutron-rich neck region connecting the two reaction partners 
survives quite a long time (around t = 500-1000 fm/c), favouring the development of
surface instabilities and mean-field fluctuations, leading to a variety of configurations
for the reaction outcome.   
Rather deformed primary PLF/TLF are observed, 
which may split, by further break-up, into 
massive fragments of comparable size.
 This effect is quite pronounced in the impact parameter window considered,
whereas for more central or more peripheral collisions, rather compact
PLF/TLF fragments 
are emerging from the reaction path.

Figure 2 shows density contour plots obtained for the same reaction at 23 MeV/A, b=5,6 fm. 
Here fragmentation times become shorter and one can observe that the neck reagion is mostly absorbed by one of the
two main fragments (PLF or TLF), inducing the formation of a rather elongated object, which may eventually breakup, 
accompanied by a fragment of more compact shape.

\subsection{Fragment recognition and PLF/TLF properties} 

As it can be noticed in Figs.1-2, multiple break-up is not actually observed over the time scales ($\approx$ 1000 fm/c) 
compatible with our dynamical description. This could be due to an overestimation of dissipative effects induced
by nucleon emission \cite{Denis_2} and/or to the approximate treatment of fluctuations in the SMF model \cite{Salvo}.  
The latter point could be cured by 
new methods to implement fluctuations in full phase space, which are presently under study \cite{BLOB}. 
A more effective fluctuating term is expected to lead to a faster dynamics, thus lowering 
energy dissipation and increasing the probability to observe a direct splitting of the system. 

However, 
the time evolution of shape observables, such as quadrupole and octupole moments, of the primary fragments emerging from 
the reaction dynamics (see Figs.1-2) can be used as an indicator of the occurrence of a subsequent break-up 
\cite{Leonid,Rizzo_2011}. 
Indeed a steadily increasing behavior of these shape observables, for a given fragment, can be associated 
with a large probability of finally observing its breakup in two pieces. 

In the SMF model, the reaction products are reconstructed by applying a coalescence procedure
to the one-body density $\rho({\bf r})$, i.e. connecting neighboring cells, in coordinate space, with density 
 $\rho({\bf r}) > \rho_0/6$. 
In this way one can also identify a ``gas'' phase   ($\rho({\bf r}) < \rho_0/6$) associated with particles that
leave rapidly the system (pre-equilibrium emission) and/or are evaporated. 
Once fragments are identified, from the knowledge of the one-body distribution function, it is possible to calculate
mass, charge, shape observables and kinematical properties.

We first focus on the behavior of quadrupole $Q_2 = \int \rho({\bf r}) (3z^2 - r^2) d{\bf r} $ 
and octupole $Q_3 = \int \rho({\bf r}) z(5z^2 - 3r^2) d{\bf r} $ moments of PLF/TLF fragments.  
  In the following, we will denote, in each event, as DF (deformed fragment) 
the one exhibiting the largest deformation, which my eventually break-up, 
and as SF(spherical fragment) the other one.   
Note that in some events both fragments can be deformed. Actually the latter situation is more likely
at the lower energy (15 MeV/A, whereas at 23 MeV/A the two fragments exhibit a different shape, 
see Figs.1-2. 

Figure 3 represents the average $Q_2$ and $Q_3$ values of the primary fragments issued from
the reaction at 23 MeV/A, b = 6 fm, as a function of time.  
Error bars reported in the figure are associated with the variance evaluated over the 100 events considered. 
In the figure, the initial time ($t_{sep}$) corresponds to the instant
when the composiste system splits into PLF and TLF fragments. 
As already suggested by the plots shown in Fig.2, one observes that, whereas the shape of one of the 
two primary fragments (the SF fragment) keeps stabilized 
around rather low $Q_2$ and $Q_3$ values, 
the deformation of the other fragment (DF), which has absorbed the neck region, increases with time, reaching a kind
of saturation around $t-t_{sep} \approx$ 300 fm/c.
\begin{figure}[t]
\includegraphics[width=9.cm]{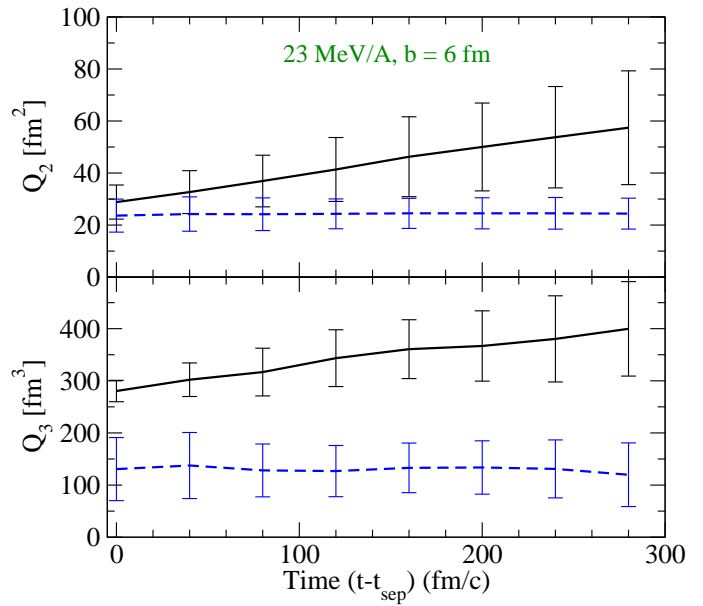}
\caption{(Color online) Time evolution of the average quadrupole (top) and octupole (bottom) moments of DF (full line) and SF (dashed line) fragments 
(see text)
obtained in  the $^{197}$Au + $^{197}$Au reaction at 23 MeV/A, b = 6 fm}
\end{figure}

We follow the reaction dynamics until $t_{stop} = 1250 (900) fm/c$ for the collisions at 15 MeV/A, b = 4(6) fm 
and $t_{stop} = 900 (750) fm/c$ at 23 MeV/A, b = 5(6) fm.
At this considered final time, PLF and TLF are well separated and the development of shape deformation is
clearly evidenced.  
Thus we assume that,
after the DF has reached its maximum degree of deformation,
it will likely split in two pieces, though we cannot provide a precise
estimation of the breakup instant and of the corresponding degree of 
alignment of the three fragments.

Figure 4 (top panel) shows the PLF-TLF distribution, evaluated at $t_{stop}$, in the plane determined by parallel and transverse velocities,
for the reaction at 23 MeV/A, b = 5-6 fm. The bottom panel displays the mass distribution of 
DF and SF fragments. It is possible to observe that the mass of the DF fragment is systematically larger,
though the effect is not so pronounced.  
Moreover, we find for the PLF/TLF fragments
an average mass of $\approx 150$ units, indicating the occurrence of a more abundant nucleon
emission (due to pre-equilibrium effects and/or evaporation) with respect to
the experimental data, reporting larger masses for PLF/TLF \cite{Isa_old,Isa}.   

\begin{figure}[t]
\includegraphics[width=8.cm]{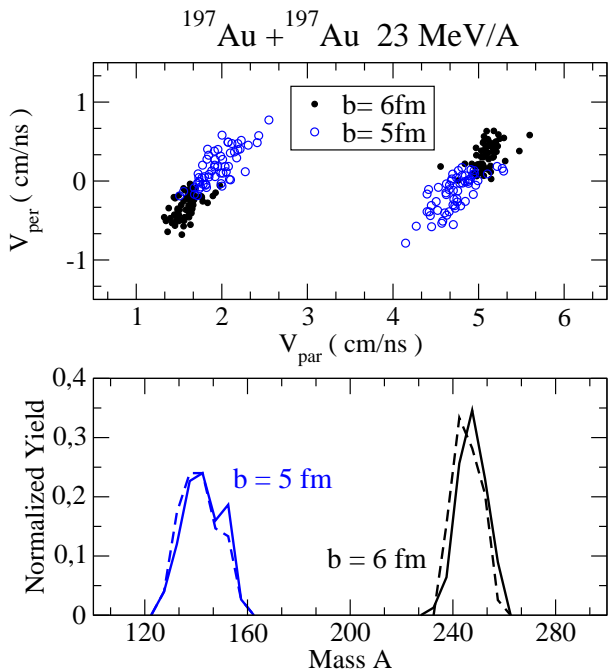}
\caption{(Color online) Top panel: Distribution in the plane determined by parallel and transverse velocities of PLF and
TLF  fragments issued from the $^{197}$Au + $^{197}$ reaction, at 23 MeV/A, b = 5-6 fm. 
Bottom panel: Mass distribution, normalized to the number of events, of DF (full line) and SF (dashed line) fragments for the same reactions
indicated above. For the reaction at b = 6 fm, the fragment masses are shifted by 100 units in the figure, 
for a better visibility.  }
\end{figure}


We then concentrate our analysis on the DF fragments and estimate their most
probable break-up configuration, as explained below and represented in Fig.5.  
\begin{figure}[t]
\includegraphics[width=7.cm]{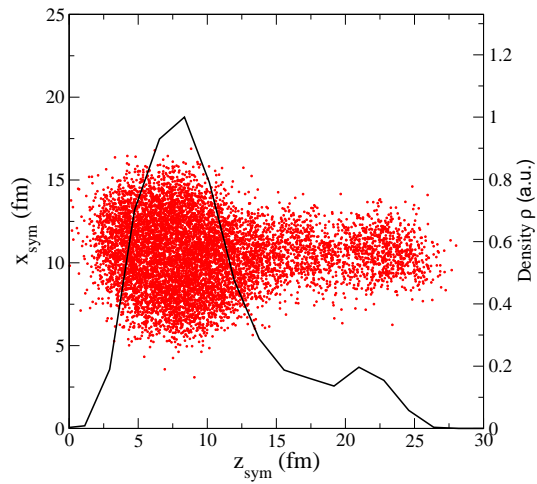} 
\caption{(Color online) Illustration of the fragment recognition procedure adopted in the calculations:
Density distribution of a DF fragment in the plane determined by its symmetry
axis and the corresponding transverse axis. The full line denotes the density profile
along the symmetry axis (in arbitrary units).}
\end{figure}
For each event,
we first evaluate the center of mass of the largest agglomerate inside the DF fragment, 
corresponding to the maximum reached by  the density profile
along its symmetry axis direction.  Then we calculate the corresponding nucleon number, 
$A2_{half}$, integrating the density from the left extreme up to the center (for a fragment orientation as shown in Fig. 5). 
The masses of the largest (A2) and smallest (A1) fragments which may originate from the break-up of the DF fragments
are evaluated as:
$A2= 2~A2_{half}$ and $A1= A_{DF} - A2$,
where $A_{DF}$ is the total mass of the primary DF fragment.  
The same procedure is followed to evaluate the charge of the two fragments.

\subsection{Fragment properties in ternary break-up}

Let us consider first the reaction at 15 MeV/A. 
In this case the neck rupture is almost symmetric between PLF and TLF, leading
to the formation of two deformed fragments in the majority of the events. 
Since we are interested in ternary events, possible breakup configurations are considered 
only for one of the two fragments at a time. Moreover,  generally speaking, 
the probability that both fragments eventually breakup, leading to quaternary events,   
should be much smaller than the probability of observing
just one rupture.  
\begin{figure}[t]
\vskip 1.0cm
\includegraphics[width=8.5cm]{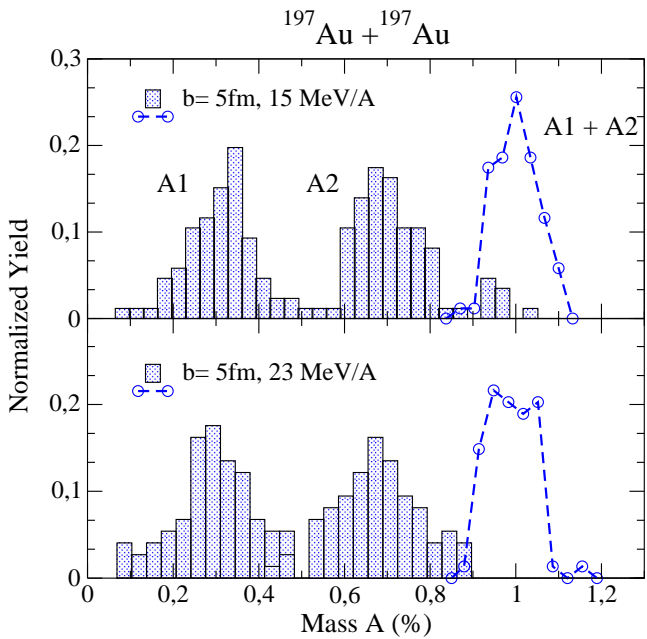}
\caption{(Color online) Mass distribution of the fragments A1 and A2, and of their sum (DF fragment),
emerging from the breakup of the DF fragment, for the reaction  $^{197}$Au + $^{197}$Au
at 15 (top) and 23 (bottom) MeV/A.  A1 denotes the smallest fragment.
Masses are normalized to the average mass of the DF fragment.}
\end{figure}

Figure 6 (top panel) 
shows the mass distribution of the lightest, A1, and heaviest, A2, fragments identified with the 
method outlined above, for the impact parameter b = 5 fm.
Very similar results are obtained for other impact parameters inside the considered window ([4-6] fm). 
Once the masses are normalized to the average mass of the DF fragments,   
results look 
close to the experimental distributions (see Ref.\cite{Isa_old,wilc}). 
In particular, the calculations are able to reproduce the distance between the peaks of the A1 and A2 
mass distributions, which amounts to about $30\%$ of the DF mass. The width of the mass distributions is
also compatible with the experimental results. 
This agreement can be considered as a nice evidence that the ternary partitioning in comparable masses comes 
from the reaction dynamics associated with semi-central impact parameters. 
However, a lack of events corresponding to symmetric ruptures
is observed in the simulations. This comes from the fact that statistical fission processes of PLF/TLF fragments are 
neglected in our analysis, which only focus on breakup mechanisms of dynamical origin.
On the other hand, this contribution is contained in the data.  

Similar results have been recently reported in the context of the improved quantum molecular dynamics (ImQMD) calculations of Ref.\cite{QMD_calc}, where a detailed comparison
with the experimental findings of Ref.\cite{Isa_old} is presented.


Results obtained at the higher bombarding energy of 23 MeV/A are shown in the bottom panel of the figure. 
As already discussed above, in addition to the partitions observed at 15 MeV/A, 
a relevant  IMF emission is also seen 
in the experimental data \cite{Cap}. In the calculations we observe, see Fig. 2, the occurrence of larger surface instabilities characterizing the
neck region, from which small fragments may also orignate.   
However, 
ternary break-up events, with fragments of comparable 
masses, are also quite likely, corresponding to configurations where   
the DF fragment is 
systematically enriched in mass (see Fig.4).
The behavior observed for the mass distribution of fragments A1 and A2 
is quite close to the one obtained at 15 MeV/A. However larger variances, reflecting the more dissipative dynamics,
are observed.  

It should be noticed that, in the calculations of Ref.\cite{QMD_calc}, 
as well as in our SMF simulations at 15 MeV/A, 
the lightest fragment (A1) emerges mainly from the neck region (see also Fig.1), thus being located at mid-velocity. 
On the other hand, in the data analysis reported in Ref.\cite{wilc}, the fragment with the largest parallel velocity (F1) has the smallest
mass. 

In the SMF simulations, this feature seems to be present at higher energy.  
In fact, as one can notice in Fig.2, at 23 MeV/A, owing to increased angular momentum effects, 
the DF fragment rotates before it reaches its maximum deformation and a subsequent break-up may take place. 
As a consequence,
in the case of a PLF break-up for instance, the lightest fragment may emerge with large positive parallel velocity. 
Indeed, at the moment of its break-up, the deformed PLF may be oriented, in coordinate space, in such a way that the matter absorbed from the
neck appears located on the right side with respect to the axis connecting TLF and PLF 
(as in the configuration of Fig.5 for instance).

Then, to establish a closer connection with the experimental analysis of Ref.\cite{wilc}, in our fragment recognition procedure
we now identify as F1 the fragment which is located on the external side of the DF fragment (i.e. on its
right(left) side in the case of the PLF(TLF) break-up), and as F2 the other one.
The corresponding mass distribution is shown in Fig.7.
\begin{figure}[t]
\includegraphics[width=8.cm]{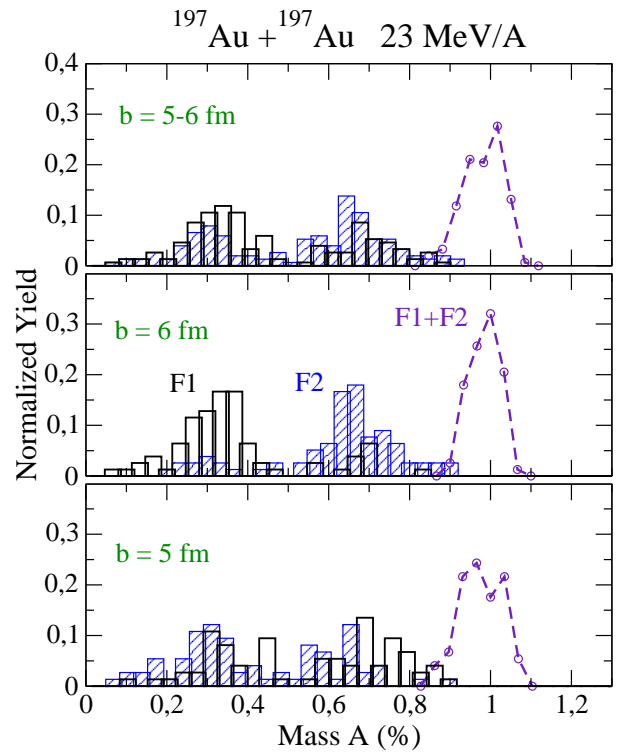}
\caption
{ (Color online) Mass distributions of fragments F1 (black histogram) and F2 (blue shaded histogram), and of their sum (dashed line), as obtained in the reaction
$^{197}$Au + $^{197}$Au at b = 5 fm (a) and b = 6 fm (b).  Panel (c) represents
the distribution corresponding to the geometrical weight of the two impact parameters.
Masses are normalized to the average mass of the DF fragment.
}
\end{figure}

The simulations show an interesting evolution, with the impact parameter, towards the features observed experimentally.
Indeed, whereas at b=5 fm the F1 fragment exhibits a wide mass distribution, at b = 6 fm it is mostly located in the
low-mass region.  These results point to the occurrence of a reaction mechanism, 
i.e. neck rupture coupled to angular momentum effects, which could explain the 
experimental observation.  
The top panel of the figure represents the distribution corresponding
to the geometrical weigth of the two impact parameters.   

As mentioned above, a significant influence of angular momentum effects on the reaction mechanism 
starts to be noticed at 23 MeV/A, whereas
in the data this effect is already present at 15 MeV/A \cite{wilc}.  This shift of beam energy may be due
to the overestimated dissipation by nucleon emission, which is a drawback of SMF 
calculations at the considered beam energies \cite{Denis_2}. Indeed the latter 
may quench angular momentum effects.  
On the other hand, in QMD-like calculations rotational effects could be missing because of the too fast 
reaction dynamics and the reduced mean-field effects \cite{Akira}. 


  




\begin{figure}[t]
\includegraphics[width=7.cm]{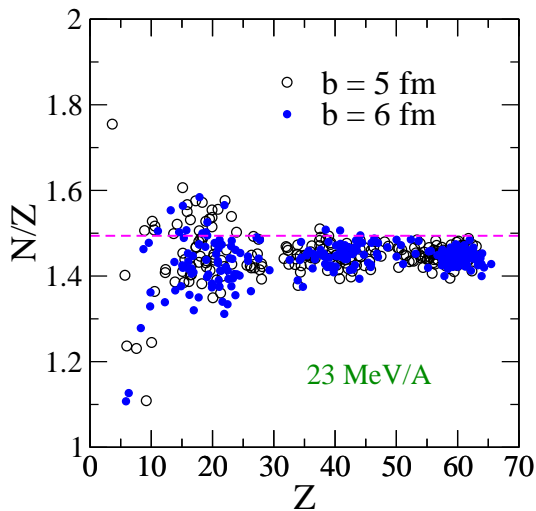}
\caption{(Color online)
N/Z ratio, as a function of the fragment charge, as obtained for the reactions at 23 MeV/A, b = 5-6 fm. 
The scattered points correspond to the simulated events. The dashed line represents the system 
initial asymmetry N/Z = 1.49. 
}
\end{figure}

Finally, we move to discuss the isotopic features of the ternary break-up events. 
Fig.8  displays the relation between the N/Z ratio and the charge number Z of the three fragments obtained, in the case of the
reaction at 23 MeV/A, b = 5,6 fm. 
As a general trend, we observe that the average N/Z ratio of the whole system at the final time, which amounts to $N/Z \approx 1.43$,  
is lower than the  
system initial asymmetry (N/Z = 1.49 for $Au$ nuclei), reflecting pre-equilibrium neutron-rich emission and/or evaporation.  
One can also see that, especially in the case of the more peripheral impact parameter (b = 6 fm), 
the lightest fragments are slightly proton-richer.  
This may be due to Coulomb polarization effects, because, as explained before, light fragments
are located, in this case, at the outer side of the system. 


\section{Conclusions}

We have untertaken an analysis of new fragmentation modes, on which recent
experimental investigations have been concentrated, 
which may develop in low-energy heavy ion collisions.
The possibility of observing ternary breakup processes of dynamical origin
is explored within the SMF model, looking at the concurrent role  
of surface mean-field instabilities, dissipative anf angular momentum  
effects.
Indeed large quadrupole and octupole deformation effects are developing 
in binary exit channels of semi-central reactions, which may lead to 
a subsequent breakup of the PLF/TLF fragments. 
The procedure based on the study of shape deformations, which was
introduced in Ref.\cite{Rizzo_2011} to evaluate fusion vs. breakup cross sections, 
is extended here to the search of multiple breakup processes.
Moreover, a fragment recognition method is introduced to identify the
most probable break-up configurations associated with deformed
PLF/TLF fragments.
For reactions at $\approx 20$ MeV/A,  SMF calculations are able to explain
the main features observed experimentally for ternary breakup events, 
namely mass partitions and the appearance of preferential emission directions. 
The model also indicates that these features emerge from a delicate balance 
between the neck dynamics and rotational effects.
Thus the analysis of these fragmentation modes allows one to get a 
deeper insight into the nature of dissipation mechanisms and the
properties of the nuclear effective interaction. 
It would be extremely interesting to extend this kind of investigations
to reactions involving neutron-rich (or even exotic) nuclei \cite{Paul}.
Indeed the reaction dynamics could be affected by the neutron-enrichment 
of the neck region, related to neutron skin effects
and/or isospin migration mechanisms \cite{rep1,Def_new}. 
One would also expect a sensitivity of the reaction mechanism,
and of the features of the emitted fragments, to the isovector terms
of the nuclear potential, opening interesting perspectives 
towards the extraction of new, independent information 
on the density behavior of the nuclear symmetry energy \cite{EPJA_book}.

\section{Acknowledgments}
This work, for V. Baran, was supported by a grant of the Romanian National
Authority for Scientific Research, CNCS - UEFISCDI, project number PN-II-ID-PCE-2011-3-0972.


\end{document}